# Non-Invasive Temporal Interference Electrical Stimulation for Spinal Cord Injury Rehabilitation: A Simulation Study

Xu Xie [a,#], Yuchen Xu [a,b,#], Huilin Mou [a], Xi Li [c], Li Zhang [d,e], Zehao Sheng [d], , Weidong Chen [a], Shaomin Zhang [a,\*], Ruidong Cheng [d,e,\*], MinminWang [a,f,\*]

[a] Qiushi Academy for Advanced Studies, Zhejiang University, Hangzhou, China.
[b] Center of Excellence in Biomedical Research on Advanced Integrated-on-Chips Neurotechnologies (CenBRAIN Neurotech), School of Engineering, Westlake University, Hangzhou, China
[c] Graduate School, Hangzhou Normal University, Hangzhou, China.
[d] Center for Rehabilitation Medicine, Department of Rehabilitation Medicine, Zhejiang Provincial People's Hospital (Affiliated People's Hospital, Hangzhou Medical College), Zhejiang Engineering Research Center for Digital-Intelligent Rehabilitation Equipment, Hangzhou, China.
[e] School of rehabilitation, Hangzhou Medical College, Hangzhou, China.
[f] Westlake Institute for Optoelectronics, Westlake University, Hangzhou, China.
[#] These authors contributed equally.

\***For Correspondence:**

Minmin Wang, minminwang@zju.edu.cn, Qiushi Academy for Advanced Studies, Zhejiang University, Hangzhou 310027, China.

**Abstract**

*Background:* Spinal cord injury (SCI) rehabilitation remains a major clinical challenge, with limited treatment options for functional recovery. Temporal interference (TI) electrical stimulation has emerged as a promising non-invasive neuromodulation technique capable of delivering deep and targeted stimulation. However, the application of TI stimulation in SCI rehabilitation remains largely unexplored.

*Methods:* This study aims to investigate the feasibility of applying non-invasive TI electrical stimulation for SCI rehabilitation. Through computational modeling, we analyzed the electric field distribution characteristics in the spinal cord under different TI stimulation configurations. Based on these findings, we propose a clinically applicable TI stimulation protocol for SCI rehabilitation..

*Results:* The results demonstrate that TI stimulation can effectively deliver focused electric fields to targeted spinal cord segments while maintaining non-invasiveness. The electric field intensity varied depending on individual anatomical differences, highlighting the need for personalized stimulation parameters. The proposed protocol provides a practical framework for applying TI stimulation in SCI rehabilitation and offers a non-invasive alternative to traditional spinal cord stimulation techniques.

*Conclusions:* This study establishes the feasibility of using non-invasive TI stimulation for SCI rehabilitation. The proposed stimulation protocol enables precise and targeted spinal cord modulation. However, further research is needed to refine personalized stimulation parameters and validate the clinical efficacy of this approach.

# 1. Introduction

Spinal cord injury (SCI) represents one of the most debilitating conditions within the field of neurology, with profound consequences for motor, sensory, and autonomic functions [1-2]. The majority of individuals with SCI experience lifelong disabilities, ranging from partial to complete loss of voluntary movement and sensation below the level of injury. As a result, the recovery and rehabilitation of SCI patients have been a major focus of research and clinical efforts [3-4]. While advancements have been made in improving the immediate survival and management of SCI, the restoration of lost functions remains a significant challenge. Conventional rehabilitation strategies typically include a combination of physical therapy, pharmacological treatments, and neuromodulation approaches, but the overall success in achieving functional recovery remains limited [5-8].

One of the most promising approaches to SCI rehabilitation involves spinal cord stimulation (SCS), which has shown potential in enhancing motor recovery and improving sensory function [9-12]. SCS can be classified into invasive and non-invasive methods. Invasive techniques, such as epidural spinal cord stimulation (eSCS), involve the implantation of electrodes directly onto the spinal cord to induce electrical impulses [13-15]. This approach has led to partial recovery of motor functions in some patients, including the ability to voluntarily control movement. However, the invasive nature of these techniques poses risks such as infection, scar tissue formation, and the need for ongoing surgical intervention. Non-invasive alternatives, such as transcutaneous electrical nerve stimulation (TENS), offer a less invasive solution by stimulating the spinal cord through the skin [16]. While TENS has been found to alleviate pain and improve some motor functions, its effectiveness for long-term rehabilitation, particularly for individuals with severe SCI, is still debated [17]. The main challenge with non-invasive approaches lies in the difficulty of targeting the deep structures of the spinal cord with sufficient precision and intensity to achieve meaningful therapeutic outcomes.

In recent years, a promising new modality known as temporal interference (TI) stimulation has emerged as a potential breakthrough in non-invasive neuromodulation techniques [18-19]. TI stimulation works by simultaneously applying two frequency electrical currents at different frequencies, with their interference producing a low-frequency modulation at the region of interest. This phenomenon enables targeted neural stimulation without the need for invasive electrodes, providing a safer and more comfortable option for patients. The key advantage of TI stimulation is its ability to focus the electric field at a specific depth and region within the nervous system, even when the stimulation is applied externally. This makes it possible to modulate deep structures, such as the spinal cord, with higher spatial precision than traditional non-invasive techniques like TENS.

Despite its promising theoretical framework, the application of temporal interference stimulation for spinal cord rehabilitation remains an area of limited research [20]. While some studies have explored the use of TI in brain-based applications, such as enhancing motor and cognitive functions [21-23], its potential for SCI rehabilitation has not been extensively investigated. The lack of research on the mechanisms of TI on spinal cord activity, as well as the absence of well-defined protocols for stimulation parameters, presents a significant obstacle for clinical translation. Moreover, the specific effects of TI on the spinal cord's neuroplasticity and its capacity to facilitate functional recovery in SCI patients are still poorly understood.

In light of these challenges, this study aims to explore the feasibility and potential of non-invasive temporal interference electrical stimulation for spinal cord rehabilitation. By leveraging advanced electrical field simulation modeling, we aim to identify the optimal stimulation parameters and determine the regions of the spinal cord that can be most effectively targeted. Furthermore, this work seeks to propose a clinically feasible TI protocol for spinal cord rehabilitation, which could serve as a basis for future experimental studies and clinical trials.

**2. Methods and materials**

In this study, EF simulations of TI are performed using Sim4Life [24], with the simulation process illustrated in Figure 1. First, all candidate transducers are placed on the realistic human model based on the location of the target area at spinal cord. Then, different combinations of two pairs of transducers are selected from the candidates for the EF simulations. Finally, the results are compared to determine the optimal transducer placement scheme.

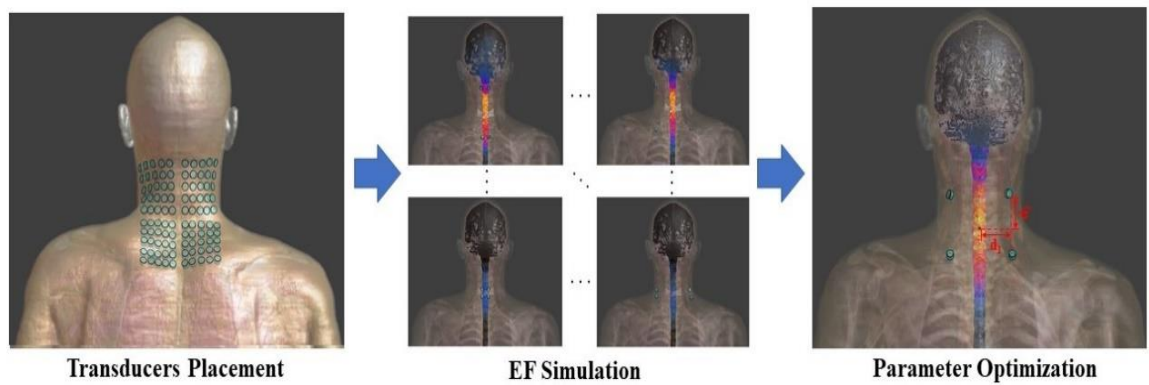

Figure 1. The overview of EF simulations of TI. The specific procedure includes: placing candidate transducers (a total of 50 pairs) based on different $d_1$ and $d_2$. According to the specific $d_1$ and $d_2$, two pairs of transducers with left-right symmetry are selected, resulting in 25 montages. The stimulation frequency for the right-side transducers pair is 1000 Hz, with the anode positioned on the top and the cathode on the bottom. For the left-side, the stimulation frequency is 1040 Hz, with the polarity opposite to that of the right side. The peak-to-peak current of the transducers is 4 mA. Finally, the simulation results of the 25 montages are compared, and the optimal montages along with its corresponding d1 and d2 values are selected.

**2.1 Realistic human model**

This study employs two realistic human models in Sim4Life for EF simulations, including a young male model, Duke (https://itis.swiss/virtual-population/virtual-population/vip3/duke/), and a young female model, Ella (https://itis.swiss/virtual-population/virtual-population/vip3/ella/). These models are high-resolution anatomical representations created based on magnetic resonance imaging (MRI) data. Each model comprises over 300 types of tissues and organs, with a spatial resolution of 0.5 × 0.5 × 0.5 mm³.

**2.2 EF simulation of TI**

Considering the stimulation frequency range of TI (kHz), the Ohmic Quasi-Static solver is used for finite element calculations in this study. The Quasi-Static Laplace equation is given by Equation (1), where σ is the electrical conductivity, and $\varphi$ is the electric potential. The EF can be obtained by $E = -\nabla\varphi$. The H-field is neglected and the E-field is calculated only in the lossy ($\sigma \neq 0$) domain.

$$\nabla \cdot \sigma \nabla \varphi = 0 \tag{1}$$

In the EF simulation of TI, the electrical conductivities of the tissues at 1 kHz involved in the Realistic Human Model are as follows [25-28]. The conductivity distribution of the tissues is assumed to be isotropic.

Table 1 The electrical conductivities of the relative tissues at 1 kHz

| Tissues | σ (S/m) | Tissues | σ (S/m) |
| --- | --- | --- | --- |
| Skin | 0.1483 | Breast | 0.0222 |
| Fat | 0.0776 | Intervertebral disc | 0.7393 |
| Muscle | 0.4610 | Artery and Vein | 0.6625 |
| Vertebra (cancellous) | 0.0805 | Lung | 0.0104 |
| Vertebra (cortical) | 0.0063 | Pancreas | 0.1450 |

| Spinal cord | 0.6110 | Liver | 0.1850 |
| Spleen | 0.1450 | Heart | 0.6625 |
| Kidney | 0.3415 | Stomach | 0.1635 |

The two pairs of transducers used in this study have stimulation frequencies of 1040 Hz and 1000 Hz, resulting in a difference frequency envelope of 40 Hz. The peak-to-peak stimulation current for both pairs of transducers is 4 mA. For each pair of transducers, Dirichlet boundary conditions are applied, with the anode transducer set to 1 V and the cathode transducer set to -1 V. The EF is then scaled to obtain the results for a current injection of 1 mA for each pair of transducers. Finally, the EF intensity generated by the two pairs of transducers are subjected to maximum modulation envelope, and the amplitude weights of the two EFs are adjusted to obtain the modulated EF distribution.

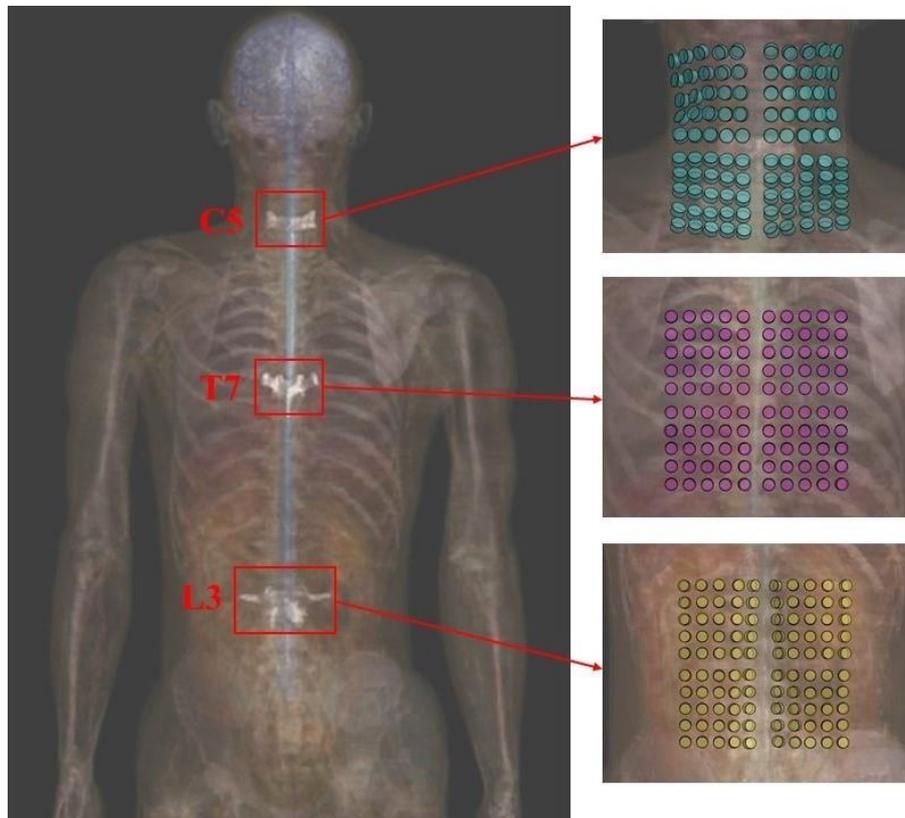

Figure 2. Three target area and corresponding 25 candidate montages. Three target area at spinal cord were selected：the cervical spinal cord at C5, the thoracic spinal cord at T7, and the lumbar spinal cord at L3. Twenty-five candidate montages were placed on the dorsal skin surface corresponding to each target area based on d1 (10 mm, 21 mm, 32 mm, 43 mm, or 54 mm) and d2 (10 mm, 25 mm, 40 mm, 55 mm, or 70 mm).

**2.3 Treatment planning**

The specific positions of the transducers are characterized by the horizontal distance $d_1$ and the vertical distance $d_2$, as shown in Figure 1. Since the spinal cord is located in the center of the body and extends vertically downward, the two pairs of transducers are symmetrically placed on either side of the spinal cord. Each pair of transducers is also symmetrically positioned above and below the target area of the spinal cord along the vertical axis. This arrangement is based both on the symmetry of human anatomical structures and the convenience for clinical applications.

Using the point on the back skin surface closest to the upper target area of the spinal cord as the origin, the four transducers forming two pairs are arranged with a horizontal distance $d_1$ and a vertical distance $d_2$ from the origin. The transducer pair on the right side of the spinal cord has a stimulation frequency of 1000 Hz, with the anode located above and the cathode below. The transducer pair on the left side of the spinal cord has a frequency of 1040 Hz, with the electrode polarity reversed compared to the right side, in order to achieve a phase difference between the currents output by the two pairs of transducers.

To investigate the effects of different montages on the target area for TI, multiple montages were configured based on different combinations of $d_1$ and $d_2$. The values of $d_1$ can be 10 mm, 21 mm, 32 mm, 43 mm or 54 mm, while the values of $d_2$ can be 10 mm, 25 mm, 40 mm, 55 mm, or 70 mm. Therefore, a total of 15 different montages were considered. In this study, three different spinal cord locations were selected as target areas: the cervical spinal cord at C5, the thoracic spinal cord at T7, and the lumbar spinal cord at L3. The optimal montage for each target area was investigated, and the patterns of the optimal montages across the three target areas were identified, as shown in Figure 2.

**2.4 Evaluation metric**

To evaluate the stimulation effects of different montages on the target areas of spinal cord, we used two evaluation metrics to evaluate the stimulation effects of different montages on the target areas of spinal cord: (a) the average EF intensity in the regions of the spinal cord wrapped by the vertebrae C5, T7, and L3. (b) The ratio of the average EF intensity of the spinal target area to the average EF intensity of the spinal cord segment (cervical, thoracic, and lumbar) where the target area is located, as shown in Equation (2).

$$\text{Focality } = \frac{average\ EF\ intensity\ at\ C5/T7/L3}{average\ EF\ intensity\ at\ Cervical/Thoracic/Lumbar\ spinal\ cord} \times 100\% \quad (2)$$

## 3. Results

### 3.1 Induced Electric Field Intensity of the TI simulation at target site

Figure 3 shows the EF simulation of TI stimulation on the cervical spinal cord C5 segment in the female model Ella using different montages. It can be observed that when $d_2$ is too low, the average EF intensity at the target area is relatively low. When $d_2 \geq 40$mm, the coverage area of high field intensity at the target area significantly increases. Additionally, when $d_2$ is fixed, if $d_1$ is too high, the average EF intensity induced by TI stimulation at the target area is also relatively low. However, in comparison, the choice of $d_2$ has a greater impact on the average EF intensity at the target area.

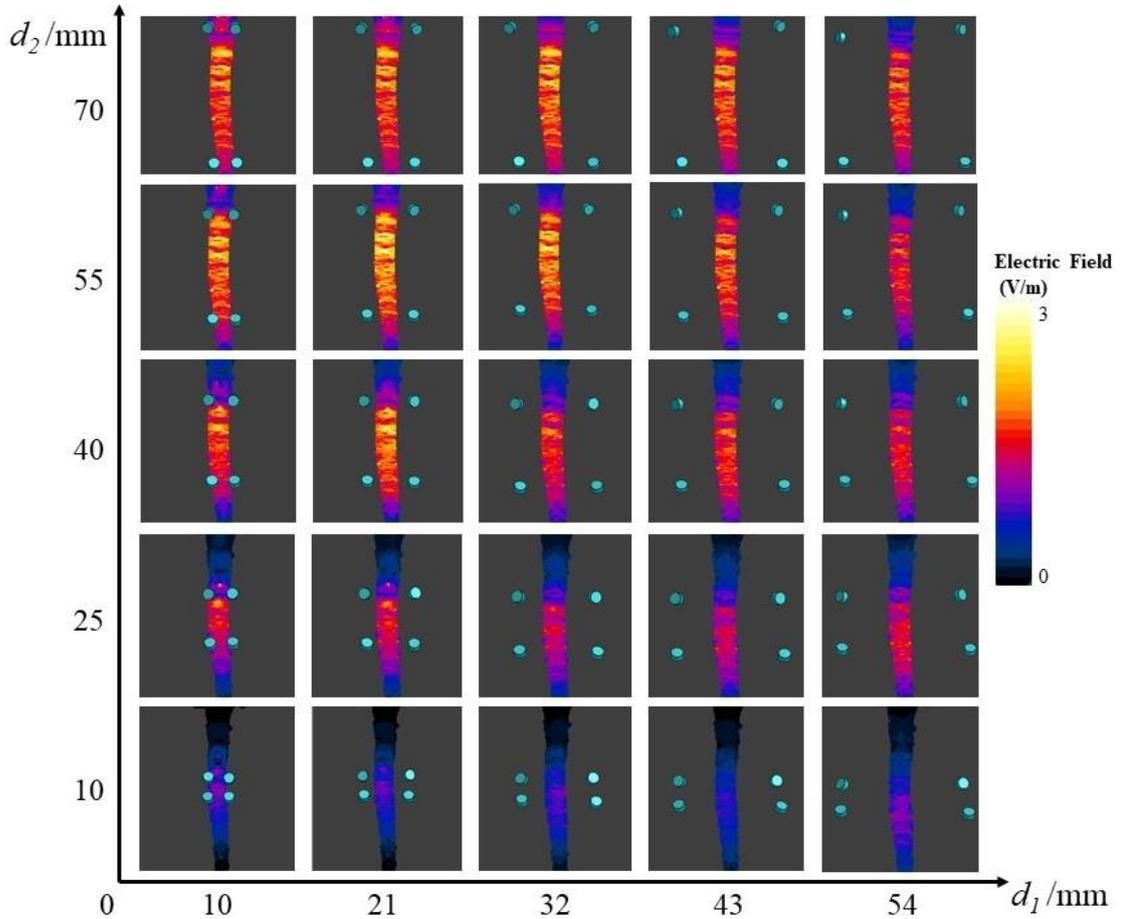
Figure 3. EF simulation of the cervical spinal cord at C5 of Ella.

Figure 4 presents the average EF intensity at the target area of three spinal cord segments (C5, T7, and L3) for TI stimulation using 25 different montages in two models (Duke and Ella). Overall, it can be observed that when $d_2$ is too small, the average EF intensity at the target area is lower. For the male model Duke, the montages that produced the maximum average EF intensity at the target areas C5, T7, and L3 had montage coordinates ($d_1$, $d_2$) of (32 mm or 43 mm, 70 mm), (10 mm, 40 mm), and (10 mm, 70 mm), with average EF intensity of 1.21 V/m, 1.01 V/m, and 0.48 V/m, respectively. The optimal value of $d_2$ for both target C5 and L3 was 70 mm, while the optimal value of $d_1$ for target T7 and L3 was 10 mm. For the female model Ella, the montages that produced the maximum average EF intensity at the target areas C5, T7, and L3 had montage coordinates ($d_1$, $d_2$) of (21 mm, 55 mm), (10 mm, 40 mm), and (10 mm, 70 mm), with average EF intensity of 1.87 V/m, 1.31 V/m, and 0.63 V/m, respectively. For both models, the optimal montages for the target T7 and L3 were the same. Notably, for the Ella model, the optimal montage for the target T7 produced a significantly higher average EF intensity than surrounding montages (at least 23.58% higher). In other cases, the neighboring montages produced average EF intensity similar to that of the optimal montage at the target area. Moreover, a comparison between the two models revealed that for all three target areas, the optimal montage in the female model Ella produced significantly higher average EF intensity than in the male model Duke (54.54%, 29.70%, and 31.25% higher for C5, T7, and L3, respectively).

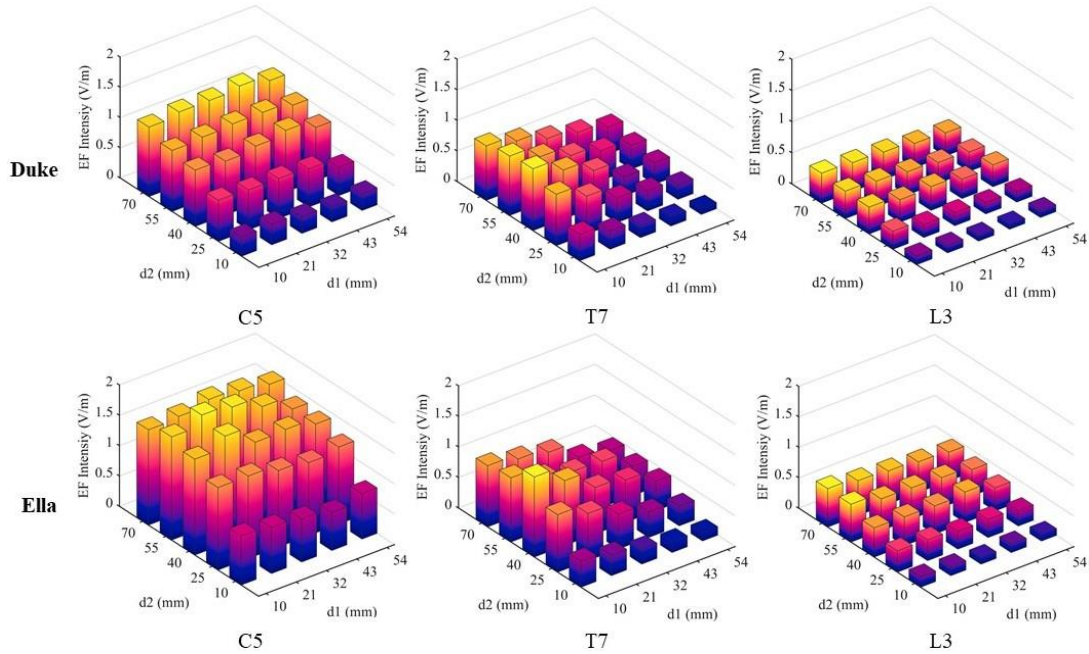

Figure 4. The intensity of the TI-induced electric field at different targets using various electrode montages.

## 3.2 Induced Electric Field focality of the TI simulation at target site

Figure 5 illustrates the EF focality for TI stimulation at three spinal cord target areas (C5, T7, and L3) using 25 different montages in two models (Duke and Ella). For the model Ella, the montage with the best focality is at coordinates (21 mm, 10 mm), while in all other cases, the montage with the best focality is located closest to the target region (10 mm, 10 mm). Since the current flows from the anode to the cathode, the spatial distribution of the current is primarily concentrated between the anode and cathode. Therefore, it is expected that when both $d_1$ and $d_2$ are low, the EF focality is better. In the male model Duke, the optimal focality for the three target regions C5, T7, and L3 is 181.36%, 431.71%, and 158.42%, respectively. For the female model Ella, the optimal focality is 221.84%, 398.72%, and 168.65%, respectively. The higher focality at the target T7 is due to the larger portion of the spinal cord occupied by the thoracic region (comprising 12 spinal segments). For all simulation conditions, when $d_1$ is fixed, the EF focality decreases as $d_2$ increases. However, when $d_2$ is fixed, the EF focality changes differently with varying $d_1$: for both models, when $d_2$ is low, an increase in $d_1$ leads to a gradual reduction in the focality at T7. For the Duke and Ella models at the target C5, the focality shows a 'U' shape and an inverted 'U' shape trend as $d_1$ increases, respectively, while the trend at the target L3 is more complex. Additionally, interestingly, when $d_2$ is large, the EF focality improves with an increase in $d_1$, although the range of change is relatively small.

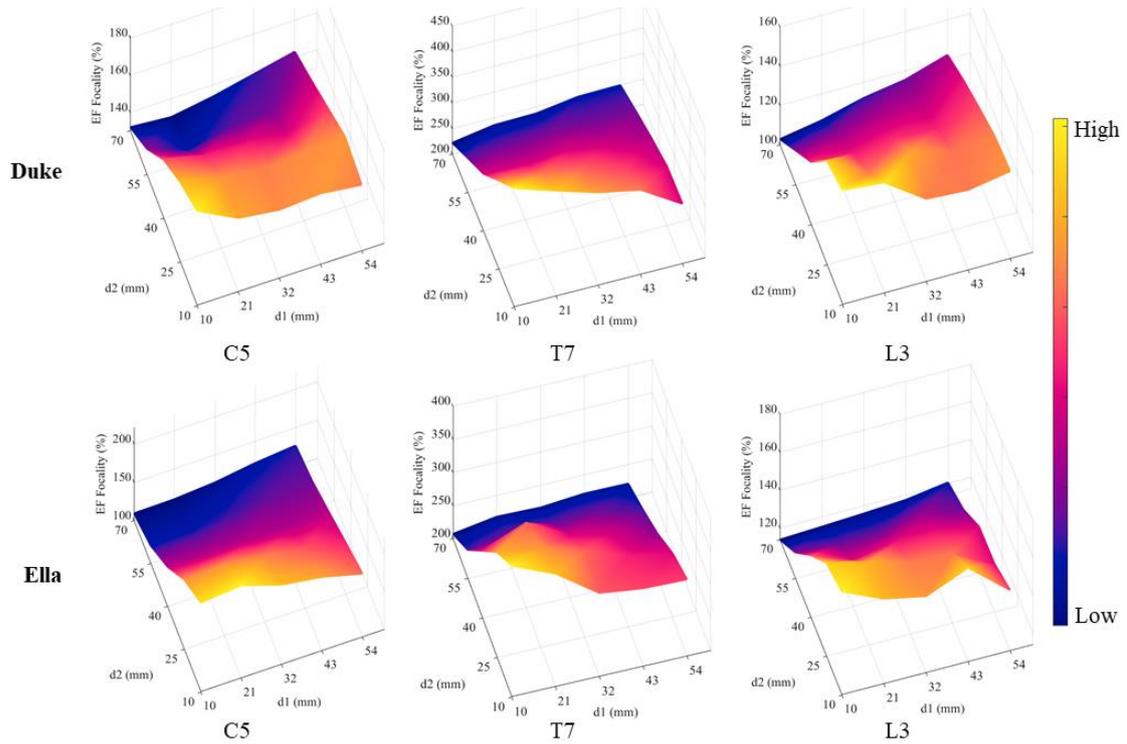
Figure 5. The focality of the TI-induced electric field at different targets using various electrode montages.

## 4. Discussion

This study presents a simulation-based framework to model the application of TI stimulation on the spinal cord, providing a detailed understanding of how electric field distribution within the body and spinal cord, which is crucial for optimizing the TI protocol and ensuring safe and effective stimulation. The proposed protocol could serve as a foundation for future clinical trials, offering a roadmap for the application of TI stimulation in SCI rehabilitation.

The protocol proposed in this study demonstrates the practical applicability of TI stimulation for SCI rehabilitation. By focusing on the C5, T7, and L3 regions of the spinal cord—corresponding to critical areas involved in motor and sensory control—the protocol aims to stimulate the regions that are most affected by SCI. This approach offers a targeted and precise neuromodulation strategy, which can focus the electric field on the desired spinal cord segments.

One of the advantages of the proposed protocol is its non-invasive nature, which allows for a more comfortable and accessible treatment for SCI patients. The ability to deliver stimulation through external electrodes placed on the skin makes this approach significantly easier to implement in clinical practice compared to invasive methods, which require surgery for electrode implantation. Additionally, the flexibility of TI stimulation, including adjustable parameters such as frequency, amplitude, and duration, enables clinicians to tailor the treatment to individual patient needs, further enhancing the potential for therapeutic success. This ease of application also opens up the possibility of at-home rehabilitation, where patients could use TI stimulation as part of a broader rehabilitation regimen, reducing the dependency on hospital visits and enabling a more continuous and personalized treatment plan. This would not only improve patient compliance but also reduce healthcare costs and resource burdens associated with long-term inpatient rehabilitation.

One critical factor that must be considered in the application is the variability in individual anatomy and tissue properties. The findings from this study underscore the importance of considering individual differences in spinal cord structure, tissue composition, and the overall electrical conductivity of tissues when designing a TI stimulation protocol. Differences in body composition, skin thickness, fat distribution, and tissue conductivity can influence how the electric field propagates and how effectively the spinal cord is modulated. For example, the thickness of the skin and fat layer could impact the penetration depth of the electrical field, particularly in areas of the body with more subcutaneous tissue. Moreover, age-related changes in tissue properties, as well as variations in individual spinal cord anatomy, could further alter the effectiveness of TI stimulation. These factors suggest that a one-size-fits-all approach to stimulation intensity may not be optimal for every patient. As such, future applications of TI stimulation in clinical practice will require a more personalized approach, where treatment intensity and duration are adjusted based on individual characteristics. The development of individualized models that account for these anatomical and physiological differences could help optimize the effectiveness of TI therapy for SCI patients, ensuring that each patient receives the most appropriate stimulation parameters for their specific condition.

While the results of this study offer promising insights into the potential of TI stimulation for SCI rehabilitation, there are several limitations that must be acknowledged. First, the modeling approach used in this study relies on only two individual body models, which may not fully capture the range of anatomical variability found in the general population. Although the models were based on realistic anatomical data, their representativeness is limited by the fact that they were not customized to account for a broader range of body types, ages, or injury levels. As such, the findings may not be fully generalizable to all SCI patients. Additionally, while this study demonstrates the theoretical feasibility of TI for SCI rehabilitation, the clinical effects of the proposed protocol need to be validated through real-world experiments. The simulations provide valuable insights into the expected outcomes of TI stimulation, but the actual therapeutic benefits, such as improvements in motor function, sensory recovery, and quality of life, can only be determined through clinical trials. The safety and long-term effects of TI stimulation in SCI patients must also be thoroughly investigated to ensure that this technique is both safe and effective for human use.

## 5. Conclusion

In conclusion, this study highlights the potential of non-invasive temporal interference stimulation as a promising tool for spinal cord injury rehabilitation. By providing a computational framework for understanding the effects of TI on the spinal cord and proposing an easily applicable stimulation protocol, the study opens new avenues for SCI treatment.

**Conflict of Interest**

All authors declare no competing interests.

**Data availability**

Data involved in this study are available upon reasonable request.

**Acknowledgments**

Research supported by the National Natural Science Foundation of China (52407261), the "Pioneer" and "Leading Goose" R&D Program of Zhejiang (2025C01137), Key Research and Development Plan of Zhejiang Province (2024C03040), Research Special Fund Project of Zhejiang Association of Rehabilitation Medicine (ZKKY2024008).

**References**

1. Lu Y, Shang Z, Zhang W, *et al*. Global incidence and characteristics of spinal cord injury since 2000-2021: a systematic review and meta-analysis. *BMC Med*. 2024;22(1):285.
2. Hu X, Xu W, Ren Y, et al. Spinal cord injury: molecular mechanisms and therapeutic interventions. *Signal Transduct Target Ther*. 2023;8(1):245.
3. Lima R, Monteiro A, Salgado AJ, Monteiro S, Silva NA. Pathophysiology and Therapeutic Approaches for Spinal Cord Injury. *Int J Mol Sci*. 2022;23(22):13833.

4. Kirshblum S, Snider B, Eren F, Guest J. Characterizing Natural Recovery after Traumatic Spinal Cord Injury. *J Neurotrauma*. 2021;38(9):1267-1284.

5. Nas K, Yazmalar L, Şah V, Aydın A, Öneş K. Rehabilitation of spinal cord injuries. *World J Orthop*. 2015;6(1):8-16.

6. Hachem LD, Ahuja CS, Fehlings MG. Assessment and management of acute spinal cord injury: From point of injury to rehabilitation. *J Spinal Cord Med*. 2017;40(6):665-675.

7. Simpson LA, Eng JJ, Hsieh JT, Wolfe DL; Spinal Cord Injury Rehabilitation Evidence Scire Research Team. The health and life priorities of individuals with spinal cord injury: a systematic review. *J Neurotrauma*. 2012;29(8):1548-1555.

8. Côté MP, Murray M, Lemay MA. Rehabilitation Strategies after Spinal Cord Injury: Inquiry into the Mechanisms of Success and Failure. *J Neurotrauma*. 2017;34(10):1841-1857.

9. Verrills P, Sinclair C, Barnard A. A review of spinal cord stimulation systems for chronic pain. *J Pain Res*. 2016;9:481-492.

10. Sdrulla AD, Guan Y, Raja SN. Spinal Cord Stimulation: Clinical Efficacy and Potential Mechanisms. *Pain Pract*. 2018;18(8):1048-1067.

11. Forouzan EJ, Rashid MY, Nasr NF, Abd-Elsayed A, Knezevic NN. The Potential of Spinal Cord Stimulation in Treating Spinal Cord Injury. *Curr Pain Headache Rep*. 2025;29(1):35.

12. de Vos CC, Meier K. Spinal cord stimulation for the treatment of chronic pain. *Nat Rev Neurol*. 2024;20(8):447-448.

13. McHugh C, Taylor C, Mockler D, Fleming N. Epidural spinal cord stimulation for motor recovery in spinal cord injury: A systematic review. *NeuroRehabilitation*. 2021;49(1):1-22.

14. Hachmann JT, Yousak A, Wallner JJ, Gad PN, Edgerton VR, Gorgey AS. Epidural spinal cord stimulation as an intervention for motor recovery after motor complete spinal cord injury. *J Neurophysiol*. 2021;126(6):1843-1859.

15. Chalif JI, Chavarro VS, Mensah E, et al. Epidural Spinal Cord Stimulation for Spinal Cord Injury in Humans: A Systematic Review. *J Clin Med*. 2024;13(4):1090.

16. Johnson MI, Paley CA, Jones G, Mulvey MR, Wittkopf PG. Efficacy and safety of transcutaneous electrical nerve stimulation (TENS) for acute and chronic pain in adults: a systematic review and meta-analysis of 381 studies (the meta-TENS study). *BMJ Open*. 2022;12(2):e051073.

17. Johnson MI. Resolving Long-Standing Uncertainty about the Clinical Efficacy of Transcutaneous Electrical Nerve Stimulation (TENS) to Relieve Pain: A Comprehensive Review of Factors Influencing Outcome. *Medicina (Kaunas)*. 2021;57(4):378.

18. Grossman N, Bono D, Dedic N, et al. Noninvasive Deep Brain Stimulation via Temporally Interfering Electric Fields. *Cell*. 2017;169(6):1029-1041.

19. Wang Y, Zeng GQ, Wang M, et al. The safety and efficacy of applying a high-current temporal interference electrical stimulation in humans. *Front Hum Neurosci*. 2024;18:1484593.

20. Sunshine MD, Cassarà AM, Neufeld E, et al. Restoration of breathing after opioid overdose and spinal cord injury using temporal interference stimulation. *Commun Biol*. 2021;4(1):107.

21. Zhu Z, Yin L. A mini-review: recent advancements in temporal interference stimulation in modulating brain function and behavior. *Front Hum Neurosci*. 2023;17:1266753.

22. Violante IR, Alania K, Cassarà AM, et al. Non-invasive temporal interference electrical stimulation of the human hippocampus [published correction appears in Nat Neurosci. 202;26(12):2252.

23. Wessel MJ, Beanato E, et al. Evidence for temporal interference (TI) stimulation effects on motor striatum. Brain Stimulation: Basic, Translational, and Clinical Research in Neuromodulation 2021;14(6): 1684.

24. Neufeld, E., Gosselin, M., Sczcerba, D., Zefferer, M., & Kuster, N. Sim4Life: A Medical Image Data Based Multiphysics Simulation Platform for Computational Life Sciences, 2012.

25. S. Gabriel, R.W. Lau, C. Gabriel, The dielectric properties of biological tissues: III. Parametric models for the dielectric spectrum of tissues, Phys. Med. Biol. 41 (11) (1996) 2271-2293.

26. Geddes LA, Baker LE. The specific resistance of biological material--a compendium of data for the biomedical engineer and physiologist. *Med Biol Eng*. 1967;5(3):271-293.

27. Balmer TW, Vesztergom S, Broekmann P, Stahel A, Büchler P. Characterization of the electrical conductivity of bone and its correlation to osseous structure. *Sci Rep*. 2018;8(1):8601.
28. Reddy GN, Saha S. Electrical and dielectric properties of wet bone as a function of frequency. *IEEE Trans Biomed Eng*. 1984;31(3):296-303.